\documentclass[journal]{IEEEtran}
\usepackage{graphics}
\usepackage{amssymb}
\usepackage{graphicx}
\usepackage{fancybox}
\usepackage{amsmath}
\usepackage{subfigure}
\usepackage{algorithm}

\usepackage[noadjust]{cite}

\usepackage{booktabs}
\usepackage{threeparttable}
\usepackage{algpseudocode}
\usepackage{amsfonts}
\usepackage{xcolor}
\usepackage{color}
\usepackage{bm}

\usepackage{tabularx}
\usepackage{multirow}

\usepackage{enumitem} %
\usepackage{hyperref}
\hypersetup{pdftex,colorlinks=true,allcolors=black}
\usepackage{makecell} %

\usepackage{stfloats} %

\definecolor{green}{RGB}{0,128,0} %

\newcommand{\mk}{\mathsf{k}}

\newcommand{\mS}{\mathsf{S}}

\newcommand{\mkstar}{\mk^{\star}}
\newcommand{\rhostar}{\rho^{\star}}

\newcommand{\dx}{\delta_x}

\newcommand{\myCRB}[1]{\mathrm{CRLB}\left\{#1\right\}}
\newcommand{\myExp}[1]{\mathbb{E}\left\{#1\right\}}

\newcommand{\myBrktRound}[1]{\left(#1\right)}
\newcommand{\myBrktSqr}[1]{\left[#1\right]}

\newcommand{\myLn}[1]{\ln\left(#1\right)}

\newcommand{\myFloor}[1]{\left\lfloor#1\right\rfloor}

\newcommand{\myEqlOverset}[1]{\overset{#1}{=}}
\newcommand{\myDef}{\overset{\triangle}{=}}
\newcommand{\myApprx}[1]{\overset{#1}{\approx}}
\newcommand{\erf}[1]{\mathrm{erf}\left(#1\right)}

\begin{document}
\title{Fast and Accurate Linear Fitting for Incompletely Sampled Gaussian Function With a Long Tail}
\author{Kai Wu, J. Andrew Zhang,~\IEEEmembership{Senior Member,~IEEE} and Y. Jay Guo,~\IEEEmembership{Fellow,~IEEE}

\thanks{K. Wu, J. A. Zhang and Y. J. Guo are with the Global Big Data Technologies Centre, University of Technology Sydney, Sydney, NSW 2007, Australia (e-mail: kai.wu@uts.edu.au; andrew.zhang@uts.edu.au; jay.guo@uts.edu.au).}%

}
\maketitle
\section{Background}
\label{secintro}

Fitting experiment data onto a curve is a common signal processing technique to extract data features and establish the relationship between variables. 
Often, we expect the curve to comply with some analytical function and then turn data fitting into estimating the unknown parameters of a function.
Among analytical functions for data fitting, the Gaussian function is the most widely used one due to its extensive applications in numerous science and engineering fields. To name just a few, Gaussian function is highly popular in statistical signal processing and analysis, thanks to the central limit theorem \cite{book_formulasTables4SP_poularikas2018handbook}; Gaussian function frequently appears in the quantum harmonic oscillator, quantum field theory, optics, lasers, and many other theories and models in Physics \cite{GF_applicationsWeb}; moreover, Gaussian function is widely applied in chemistry for depicting molecular orbitals, in computer science for imaging processing and in artificial intelligence for defining neural networks.

Fitting a Gaussian function, or simply Gaussian fitting, is consistently of high interest to the signal processing community \cite{GF_SPletter2013,GF_caruana1986fast,GF_Guo_SPM2011,GF_Nahhal_SPM2019}. 
Since the Gaussian function is underlain by an exponential function, it is non-linear and not easy to be fitted directly.
One effective way of counteracting its exponential nature is to apply the natural logarithm, which has been applied in transferring the Gaussian fitting into a linear fitting \cite{GF_caruana1986fast}. However, the problem of the logarithmic transformation is that it makes the noise power vary over data samples, which can result in biased Gaussian fitting. 
The weighted least square (WLS) fitting is known to be effective in handling uneven noise backgrounds \cite{book_kay1993fundamentals_estimation}. However, as unveiled in \cite{GF_Guo_SPM2011}, the ideal weighting for linear Gaussian fitting is directly related to the unknown Gaussian function. 
To this end, an iterative WLS is developed in \cite{GF_Guo_SPM2011}, starting with using the data samples (which are noisy values of a Gaussian function) as weights and then iteratively reconstructing the weights using the previously estimated function parameters.

For the iterative WLS, the number of iterations required for a satisfactory fitting performance can be large, particularly when an incompletely sampled Gaussian function with a long tail is given (see Fig. \ref{fig: WLS fitted Gaussian curves}(a) for such a case). Establishing a good initialization is a common strategy for improving the convergence speed and performance of an iterative algorithm. \textit{Noticing the unavailability of a proper initialization for the iterative WLS, 
we aim to fill the blank by developing a high-quality one in this article.} 
To do so, we introduce a few signal processing tricks to develop high-performance initial estimators for the three parameters of a Gaussian function. Applying our initial fitting results, not only the efficiency of the iterative WLS is substantially improved, its accuracy is also greatly enhanced, particularly for noisy and incompletely sampled Gaussian functions with long tails. These will be demonstrated by simulation results.

\section{Prior Art and Motivation}
\label{sectiptrick}

Let us start by elaborating on the signal model. A Gaussian function can be written as 
\begin{align}\label{eq: f(x)}
	f(x) = Ae^{-\frac{(x-\mu)^2}{2\sigma^2}},
\end{align}
where $ x $ is the function variable, and $ A $, $ \mu $ and $ \sigma $ are the parameters to be estimated. They represent the height, location and width of the function, respectively. 
Directly fitting $ f(x) $ can be cumbersome due to the exponential function. 
A well-known opponent of exponential is the natural logarithm. 
Indeed, taking the natural logarithm of both sides of (\ref{eq: f(x)}), we can obtain the following polynomial after some basic rearrangements,
\begin{align} \label{eq: ln(f(x))}
	\ln\myBrktRound{f(x)} = a + bx + cx^2, 
\end{align}
where the coefficients $ a,b $ and $ c $ are related to the Gaussian function parameters $ \mu $, $ \sigma $ and $ A $. Based on (\ref{eq: f(x)}) and (\ref{eq: ln(f(x))}), it is easy to obtain
\begin{align} \label{eq: mu sigma and A wrt a b and c}
	\mu = \frac{-b}{2c};~\sigma=\sqrt{\frac{-1}{2c}};~A=e^{a-b^2/(4c)}.
\end{align}
We see that the estimations of $ \mu $, $ \sigma $ and $ A $
can be done through estimating $ a $, $ b $ and $ c $.
Since
$ a $, $ b $ and $ c $ are coefficients of a polynomial, 
they can be readily estimated by employing linear fitting based on e.g., the least square (LS) criterion \cite{book_kay1993fundamentals_estimation}.

In modern signal processing, we generally deal with noisy digital signals. Thus, instead of $ f(x) $ given in (\ref{eq: f(x)}), the following signal is more likely to be dealt with, 
\begin{align}\label{eq: y[n]}
	y[n] = f[n] + \xi[n],~\mathrm{s.t.}~f[n] = f(n\dx),~n=0,\cdots,N-1,
\end{align}
where $ n $ is the sample index, $ \dx $ is the sampling interval of $ x $, and $ \xi[n] $ is an additive white Gaussian noise. 
If we take the natural logarithm of $ y[n] $, we then have
\begin{align} \label{eq: ln (y[n])}
	& \myLn{y[n]} = \myLn{f[n] + \xi[n]} = \myLn{f[n]\myBrktRound{1 + \frac{\xi[n]}{f[n]}}} , \nonumber\\
	& \approx \myLn{f[n]} + \frac{\xi[n]}{f[n]} = a + b\dx n + c\dx^2 n^2 + \frac{\xi[n]}{f[n]},
\end{align}
where the first-order Taylor series $ \ln(1+x)\approx x $ is applied to get the approximation, and $ \myLn{f[n]} $ is written into a polynomial form based on (\ref{eq: ln(f(x))}). Next, we review several linear fitting methods, through which the motivation of this work will be highlighted. 

For the sake of conciseness, we employ vector/matrix forms for the sequential illustrations. 
In particular, let us
define the following three vectors, 
\begin{align} \label{eq: y x theta bf}
	\begin{array}{c}
		\mathbf{y} = \myBrktSqr{ \myLn{y[0]},\myLn{y[1]},\cdots,\myLn{y[N-1]} }^{\mathrm{T}}, \\
		\mathbf{x} = \myBrktSqr{ {0},{\dx},\cdots,(N-1)\dx }^{\mathrm{T}},\\
		\bm{\theta} = [a,b,c]^{\mathrm{T}}.
	\end{array}
\end{align}
Then, based on (\ref{eq: ln (y[n])}), the linear Gaussian fitting problem can be conveniently written as
\begin{align}\label{eq: y bf LS problem}
	\mathbf{y} = \mathbf{X}\bm{\theta} + \bm{\xi}, ~\mathrm{s.t.}~\mathbf{X} = \myBrktSqr{\bm{1},\mathbf{x},\mathbf{x}\odot\mathbf{x}},
\end{align}
{where $ \bm{\xi} $ denotes a column vector stacking the noise terms $ \frac{\xi[n]}{f[n]}~(n=0,1,\cdots,N-1) $ in (\ref{eq: ln (y[n])}), and  $ \odot $ denotes the point-wise product.} {Due to the use of the above vector/matrix forms, the estimators reviewed below look different from their descriptions in the original work. However, regardless of the forms, they are the same in essence.}

\subsection{Least Square (LS) Fitting}

 The first fitting method \cite{GF_caruana1986fast} reviewed here applies LS on (\ref{eq: y bf LS problem}) to estimate the three unknown coefficients in $ \bm{\theta} $. 
 The solution is classical and can be written as \cite{book_kay1993fundamentals_estimation}
\begin{align} \label{eq: y=X theta vector form}
	\hat{\bm{\theta}} = \mathbf{X}^{\dagger} \mathbf{y} = \myBrktRound{\mathbf{X}^{\mathrm{T}}\mathbf{X}}^{-1}\mathbf{X}^{\mathrm{T}}\mathbf{y},
\end{align}
where $ \mathbf{X}^{\dagger} $ denotes the pseudo-inverse of $ \mathbf{X} $. 
The LS fitting is simple but not without problems. As can be seen from (\ref{eq: ln (y[n])}), the additive white Gaussian noise $ \xi[n] $ is divided by $ f[n] $. The division can severely increase the noise power at $ n $'s with $ f[n]\approx 0 $, causing the noise enhancement problem. As a consequence of the problem, 
LS can suffer from poor fitting performance, particularly when the majority of samples are from {the tail region of a Gaussian function}, such as the one plotted in Fig. \ref{fig: WLS fitted Gaussian curves}(a).

\subsection{Iterative Weight LS (WLS) Fitting}\label{subsec: iterative WLS}

To solve the noise enhancement issue, the authors of \cite{GF_Guo_SPM2011} propose to replace the LS with WLS. In contrast to LS, which treats each sample equally, WLS applies different weights over samples. The purpose of weighting is to counterbalance noise variations. For this purpose, $ f[n] $ is the ideal weight, as seen from 
(\ref{eq: ln (y[n])}). However, $ f[n] $ is the digital sample of an unknown Gaussian function. 

To solve the problem, an iterative WLS is developed in \cite{GF_Guo_SPM2011}, where $ y[n] $ (the noisy version of $ f[n] $) is used as the initial weight. From the second iteration, the weight is constructed using the previous estimates of $ a $, $ b $ and $ c $ based on the relation depicted in (\ref{eq: ln(f(x))}). Let $ \mathbf{w}_i $ denote the weighting vector at the $ i $-th iteration, collecting the weights over $ n=0,1,\cdots,N-1 $. Then, the iterative WLS can be executed as
\begin{align}\label{eq: WLS}
	& \hat{\bm{\theta}}_i = \myBrktRound{\mathbf{X}_i^{\mathrm{T}}\mathbf{X}_i}^{-1}\mathbf{X}_i^{\mathrm{T}}\mathbf{y}_i,~i=0,1,\cdots\nonumber\\
	\mathrm{s.t.}~& ~\mathbf{y}_i = \mathbf{w}_i\odot \mathbf{y};~\mathbf{X}_i = \myBrktSqr{\mathbf{w}_i,\mathbf{w}_i\odot\mathbf{x},\mathbf{w}_i\odot\mathbf{x}\odot\mathbf{x}} \nonumber\\
	& ~ \mathbf{w}_i = e^{\mathbf{y}}~\mathrm{if}~i=0;\mathrm{~otherwise},~\mathbf{w}_i = e^{\mathbf{X}\hat{\bm{\theta}}_{i-1}},
\end{align}
where $ \mathbf{x} $ and $ \mathbf{y} $ are given in (\ref{eq: y x theta bf}) and $ \mathbf{X} $ in (\ref{eq: y bf LS problem}). The exponential is calculated point-wise in the last row.

Provided a sufficiently large number of iterations, the iterative WLS can achieve a high-performance Gaussian fitting, generally better than LS. However, the more iterations, the more time-consuming the iterative WLS would be. Moreover, the required number of iterations to achieve the same fitting performance changes with the proportion of the tail region. 

\begin{figure}[t!]
	\centering
	\includegraphics[width=8.5cm]{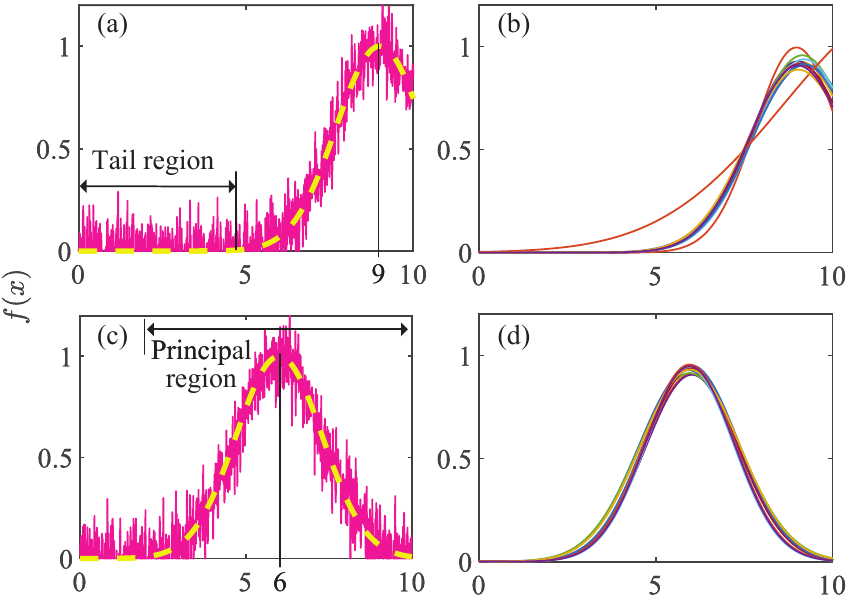}
	
	\vspace{-1mm}
	
	$ ~~~~~ x $
	
	\caption{The noisy samples of the Gaussian function with $ A=1 $, $ \mu=9 $ and $ \sigma=1.3 $ are plotted in Fig. \ref{fig: WLS fitted Gaussian curves}(a), where the dash curve is the function without noise. 
	The iterative WLS, as illustrated in (\ref{eq: WLS}), is run $ 10 $ times, each time with $ 12 $ iterations and independently generated noise, leading to the fitting results in Fig. \ref{fig: WLS fitted Gaussian curves}(b). Other than resetting $ \mu=6 $, Figs. \ref{fig: WLS fitted Gaussian curves}(c) and \ref{fig: WLS fitted Gaussian curves}(d) plot the same results as in Figs. \ref{fig: WLS fitted Gaussian curves}(a) and \ref{fig: WLS fitted Gaussian curves}(b), respectively. As shown in the figure, the range of $ x $ is $ [0,10] $, where the sampling interval is set as $ \dx=0.01 $.}
	\label{fig: WLS fitted Gaussian curves}
\end{figure}

Take the Gaussian function in Fig. \ref{fig: WLS fitted Gaussian curves}(a) for an illustration. The tail region is about half the whole sampled region. Perform $ 12 $ numbers of iterations based on (\ref{eq: WLS}) for $ 10 $ trials, each adding independently generated noise onto the same Gaussian function with $ A=1 $, $ \mu=9 $ and $ \sigma=1.3 $. The fitting results are given in Fig. \ref{fig: WLS fitted Gaussian curves}(b). We see that some fitting results still substantially differ from the true function, even after $ 12 $ iterations. In contrast, perform the same fitting as above but change $ \mu$ to $6 $ (equivalently reducing tail region).
The fitting results of $ 10 $ independent trials are plotted in Fig. \ref{fig: WLS fitted Gaussian curves}(d). Obviously, the results look much better than those in Fig. \ref{fig: WLS fitted Gaussian curves}(b). 

A common solution to
reducing the number of iterations required by an iterative algorithm is a good initialization. In our case, the quality of the initial weight vector, i.e., $ \mathbf{w}_0 $ in (\ref{eq: WLS}) can affect the overall number of iterations required by the iterative WLS to converge. Moreover, if the way of initializing $ \mathbf{w}_0 $ can be immune to the proportion of the tail region, the convergence performance of the iterative WLS can then be less dependent on the proportion of the tail region.
\textit{Our work is mainly aimed at designing a way of initializing the weight vector of the iterative WLS so as to reduce the number of overall iterations and relieve the dependence of fitting performance on the proportion of the tail region.}

\subsection{An Interesting but Not-Good-Enough Initialization}
\label{subsec: Nahhal's sigma estimation}

An initialization for iterative WLS-based linear Gaussian fitting is developed in \cite{GF_Nahhal_SPM2019}, which was originally motivated by separately fitting the parameters of a Gaussian function. In particular, exploiting the following relation
\begin{align} \label{eq: integral f(x)}
	\int_{-\infty}^{\infty} f(x)~\mathrm{d}x = A\sqrt{2\pi}\sigma. 
\end{align}  
the work proposes a simple estimation of $ \sigma $, as given by
\begin{align} \label{eq: sigma estimate nahhal}
	\hat{\sigma} = \frac{1}{\hat{A}}\sum_{n=0}^{N-1}y[n] \dx,
\end{align} 
where $ y[n] $ is the noisy sample of the Gaussian function to be estimated, $ \dx $ is the sampling interval of $ x $, and the summation approximates the above integral of $ f(x) $. Moreover, $ \hat{A} $ is estimated by
\begin{align} \label{eq: max y[n] for A and n estimates}
	\myBrktSqr{\hat{A},\hat{n}}=\max_n~y[n].
\end{align}
Similar to how $ \mathrm{max}(\cdot) $ works in MATLAB \cite{Matlab_helpCenter}, $ \hat{n} $ is the index where the maximization is achieved. 
According to
the middle relation in (\ref{eq: mu sigma and A wrt a b and c}), $ c $ can be estimated as $ \hat{c}=\frac{-1}{2\hat{\sigma}^2} $. Thus, the work \cite{GF_Nahhal_SPM2019} removes $ c $ from the parameter vector $ \bm{\theta} $ given in (\ref{eq: y x theta bf}) and employs an iterative WLS, similar to (\ref{eq: WLS}) yet with reduced dimension, to estimate $ a $ and $ b $. 

A major error source of $ \hat{\sigma} $ obtained in (\ref{eq: sigma estimate nahhal}) is the approximation error from using the summation in (\ref{eq: sigma estimate nahhal})
to approximate the integral given in (\ref{eq: integral f(x)}). Even if $ \dx $ is fine enough, the approximation can still be problematic, depending on the proportion of the tail region in the sampled Gaussian function. For example, if the sampled function has a shape similar to the one given in Fig. \ref{fig: WLS fitted Gaussian curves}(c), we know that the summation can well approximate the integral given a fine $ \dx $. However, if the sampled function has a shape like the curve in Fig. \ref{fig: WLS fitted Gaussian curves}(a), the approximation error will be large regardless of $ \dx $. A condition is given in \cite{GF_Nahhal_SPM2019}, stating when the summation in (\ref{eq: sigma estimate nahhal}) can well approximate the integral in (\ref{eq: integral f(x)}). Nevertheless, \textit{what shall we do when the condition is not satisfied}, which can be inevitable in practice, is yet to be answered.

\section{Proposed Gaussian Fitting}

Looking at Fig. \ref{fig: WLS fitted Gaussian curves}(a), we know the summation in (\ref{eq: sigma estimate nahhal}) cannot approximate the integral in (\ref{eq: integral f(x)}). Now, instead of using the summation to approximate something unachievable, how about looking into a different question: ``\textit{What can be approximated using the summation given in (\ref{eq: sigma estimate nahhal})?}'' We answer this question by performing the following computations (they look complex but are easily understandable),
\begin{align} \label{eq: sum y[n]dx ... }
	& \sum_{n=0}^{N-1}y[n] \dx\myApprx{(a)}
	\sum_{n=0}^{N-1}\dx f[n] 
	\myApprx{(b)} \int_{0}^{N\dx} f(x)~\mathrm{d}x \\
	& \myEqlOverset{(c)} \int_{0}^{\mu} f(x)~\mathrm{d}x + \int_{\mu}^{N\dx} f(x)~\mathrm{d}x\nonumber\\
	& \myEqlOverset{(d)} \frac{1}{2} \int_{0}^{2\mu} f(x)~\mathrm{d}x +\frac{1}{2} \int_{\mu-(N\dx-\mu)}^{N\dx} f(x)~\mathrm{d}x \nonumber\\
	& \myEqlOverset{(e)} \frac{\sqrt{2\pi}A\sigma}{2} ~ \erf{\frac{\mu}{\sigma\sqrt{2}}} + \frac{\sqrt{2\pi}A\sigma}{2} ~ \erf{\frac{N\dx-\mu}{\sigma\sqrt{2}}}, \nonumber
\end{align} 
where $ \erf{\cdot} $ is the so-called error function. It can be defined as \cite{book_formulasTables4SP_poularikas2018handbook}
\begin{align}\label{eq: erf function}
	\erf{z} = \frac{2}{\sqrt{\pi}} \int_{0}^{z} e^{-t^2}~\mathrm{d}t.
\end{align}
How each step in (\ref{eq: sum y[n]dx ... }) is obtained is detailed below:
\begin{enumerate}
	\item[$ \myApprx{(a)} $:] This step replaces $ y[n] $ with $ f[n] $ by omitting the noise term $ \xi[n] $, as given in (\ref{eq: y[n]});
	
	\item[$ \myApprx{(b)} $:] This is how integral is often introduced in a math textbook. While the left side of $ \myApprx{(b)} $ approximately calculates the area below $ f(x) $, the right side does so exactly;
	
	\item[$ \myEqlOverset{(c)} $:] The integrating interval is split into contiguous halves;
	
	\item[$ \myEqlOverset{(d)} $:] The integrating interval of each integral in $  \myEqlOverset{(c)} $ is doubled in such a way that it becomes symmetric against $ x=\mu $. Since $ f(x) $ is also symmetric against $ x=\mu $; see (\ref{eq: f(x)}), the scaling coefficient $ \frac{1}{2} $ can counterbalance the extension of integrating interval;
	
	\item[$ \myEqlOverset{(e)} $:] It is based on a known fact \cite{book_formulasTables4SP_poularikas2018handbook} 
	\[\int_{\mu-\epsilon}^{\mu+\epsilon} f(x) \mathrm{d}x = \sqrt{2\pi}A\sigma ~ \erf{\frac{\epsilon}{\sigma\sqrt{2}}}.\]
	
\end{enumerate}

Similar to $ \myEqlOverset{(c)} $ in (\ref{eq: sum y[n]dx ... }), we can also split the summation $ \sum_{n=0}^{N-1}y[n] \dx $. Doing so, the two integrals on the right-hand side of $ \myEqlOverset{(c)} $ in (\ref{eq: sum y[n]dx ... }) can be, respectively, approximated by 
\begin{align} \label{eq: S alpha S beta}
	\mS_{\beta} \myDef \sum_{n=0}^{\hat{n}-1}y[n] \dx~~\mathrm{and}~~\mS_{\alpha} \myDef  \sum_{n=\hat{n}}^{N-1}y[n] \dx.
\end{align}
where $ \hat{n} $ is obtained in (\ref{eq: max y[n] for A and n estimates}). (Note that $ \hat{n}\dx $ is an estimate of $ \mu $.) Moreover, tracking the computations in (\ref{eq: sum y[n]dx ... }), we can easily attain
\begin{align} \label{eq: S alpha = erf...; S beta = erf...}
	\begin{gathered}
		\mS_{\beta} 
		\approx \frac{\sqrt{2\pi}\hat{A}\sigma}{2} ~ \erf{\frac{\hat{n}\dx}{\sigma\sqrt{2}}} ;\\
	\mS_{\alpha}
	\approx \frac{\sqrt{2\pi}\hat{A}\sigma}{2} ~ \erf{\frac{N\dx-\hat{n}\dx}{\sigma\sqrt{2}}},
	\end{gathered}
\end{align}
where $ A $ and $ \mu $ have been replaced by their estimates given in (\ref{eq: max y[n] for A and n estimates}).
The two equations in (\ref{eq: S alpha = erf...; S beta = erf...}) provide possibilities for estimating $ \sigma $ that is the only unknown left. However, due to the presence of the non-elementary function $ \erf{\cdot} $, analytically solving $ \sigma $ from the equations is non-trivial. 
Moreover, we have two equations but one unknown. How to constructively exploit the information provided by both equations is also a critical problem. 
Below, we first develop an efficient method to estimate $ \sigma $ from either equation in (\ref{eq: S alpha = erf...; S beta = erf...}), resulting in two estimates of $ \sigma $; we then derive an asymptotically optimal combination of the two estimates. 

\subsection{Efficient Estimation of $ \sigma $} \label{subsec: sigma estimation}

The two equations in (\ref{eq: S alpha = erf...; S beta = erf...}) have the same structure. \textit{So let us focus on the top one for now.} While solving $ \sigma $ analytically is difficult, numerical means can be resorted to. 
As commonly done, we can select a large region of $ \sigma $, discretize the region into fine grids, evaluate the values of the right-hand side in (\ref{eq: S alpha = erf...; S beta = erf...}) on the grids, and identify the grid that leads to the closest result to $ \mS_{\beta} $. 

The above steps are regular but not practically efficient. This is because evaluating the right-hand side of the equation in (\ref{eq: S alpha = erf...; S beta = erf...}) needs the calculation of $ \erf{\cdot} $ for each $ \sigma $-grid. From (\ref{eq: erf function}), we see that $ \erf{\cdot} $ itself is an integral result. If we calculate $ \erf{\cdot} $ on-board, it would be approximated by a summation over sufficiently fine grids of the integrating variable. This can be highly time-consuming, particular given that $ \erf{\cdot} $ needs to be calculated for each entry in a large set of $ \sigma $-grids. Alternatively, we may choose to store a look-up table of $ \erf{\cdot} $ on board. This is doable but can also be troublesome, for the reason that the parameter of $ \erf{\cdot} $, as dependent on the parameters of the Gaussian function to be fitted, can span over a large range in different applications. 
The trouble, however, can be relieved through a simple variable substitution. 

Making the substitution of $ \hat{n}\dx=\mk \sigma $ in (\ref{eq: S alpha = erf...; S beta = erf...}), we obtain
\begin{align} \label{eq: S beta appr erf mk...}
	\mS_{\beta} 
	\approx \frac{\sqrt{2\pi}\hat{A}\hat{n}\dx}{2\mk} ~ \erf{\frac{\mk}{\sqrt{2}}}.
\end{align}
{Clearly, the dependence of the $ \erf{\cdot} $ function on $ \mu $ (represented by $ \hat{n}\dx $) and $ \sigma $, as shown in (\ref{eq: S alpha = erf...; S beta = erf...}), is now removed, making the $ \erf{\cdot} $ function solely related to the coefficient $ \mk $.} Therefore, a significance of the variable substitution is that 
one look-up table of $ \erf{\cdot} $ can be applied to a variety of applications with different Gaussian function parameters. Assuming $ k=\mkstar $ makes the right-hand side of (\ref{eq: S beta appr erf mk...}) closest to $ \mS_{\beta} $, $ \sigma $ can then estimated as $ \frac{\hat{n}\dx}{\mkstar} $. Similarly, we can make the substitution for the bottom equation in (\ref{eq: S alpha = erf...; S beta = erf...}) and obtain another estimate of $ \sigma $.
In summary, the two estimators can be established as follows,
\begin{align} \label{eq: sigma alpha sigma beta}
	\begin{gathered}
		\hat{\sigma}_{\alpha} \substack{= \frac{(N-\hat{n})\dx}{\mkstar}}\substack{;~\mathrm{s.t.}~
		\mkstar:\arg \min_{\mk} \myBrktRound{\mS_{\alpha} - \frac{\sqrt{2\pi}\hat{A}(N-\hat{n})\dx}{2\mk} ~ \erf{\frac{\mk}{\sqrt{2}}}}}^2; \\
		\hat{\sigma}_{\beta} \substack{= \frac{\hat{n}\dx}{\mkstar}}\substack{;~\mathrm{s.t.}~
		\mkstar:\arg \min_{\mk} \myBrktRound{\mS_{\beta} - \frac{\sqrt{2\pi}\hat{A}\hat{n}\dx}{2\mk} ~ \erf{\frac{\mk}{\sqrt{2}}}}}^2.
	\end{gathered}
\end{align}
The two estimates would have different qualities, depending on how many samples are used for each. 
This further suggests that combining them is not as trivial as simply averaging them. Next, we develop a constructive way of combining them.

\subsection{Asymptotically Optimal $ \sigma $ Estimation} \label{subsec: combining two estimates}

The two estimates obtained in (\ref{eq: sigma alpha sigma beta}) are mainly differentiated by how many samples are used in their estimations. 
Therefore, identifying the impact of the employed samples on the estimation performance is helpful in determining a way to combine the estimates. 
One of the most common performance metrics for an estimator is the Cram\'er-Rao lower bound (CRLB) \cite{book_kay1993fundamentals_estimation}. 
This points the direction of our next move. 

Let us check the CRLB of $ \hat{\sigma}_{\beta} $ first. It is estimated using the samples $ y[n] $ for $ n=0,1,\cdots,\hat{n}-1 $. Referring to (\ref{eq: y[n]}), $ y[0], y[1],\cdots,y[\hat{n}-1] $ are jointly normally distributed with different means but the same variance, as given by $ \sigma_{\xi}^2 $. Recall that $ \sigma_{\xi}^2 $ is the power of the noise term $ \xi[n]~(\forall n) $ given in (\ref{eq: y[n]}). Since we focus here on investigating the estimation performance of $ \sigma $, we assume that $ A $ and $ \mu $ are known. Then, the CRLB of $ \hat{\sigma}_{\beta} $ estimation, as obtained based on $ y[0], y[1],\cdots,y[\hat{n}-1] $, can be computed by 
\begin{align} \label{eq: CRLB sigma beta}
	\myCRB{\hat{\sigma}_{\beta}} = \frac{\sigma_{\xi}^2}{\sum_{n=0}^{\hat{n}-1} \left( \frac{\partial f[n]}{\partial \sigma } \right)^2 } = \frac{\sigma_{\xi}^2}{\frac{\sum_{n=0}^{\hat{n}-1}f[n]^2(\mu-\dx n)^4	}{\sigma^6} },
\end{align}
where the middle result is a simple application of \cite[Eq.(3.14)]{book_kay1993fundamentals_estimation}, 
and the first partial derivative of $ f[n] $ can be readily derived based on its expression given in (\ref{eq: y[n]}). 
With reference to (\ref{eq: CRLB sigma beta}), we can directly write the CRLB of $ \hat{\sigma}_{\alpha} $, as given by 
\begin{align} \label{eq: CRLB sigma alpha}
	\myCRB{\hat{\sigma}_{\alpha}} = \frac{\sigma_{\xi}^2}{\frac{\sum_{\hat{n}}^{N-1}f[n]^2(\mu-\dx n)^4	}{\sigma^6} },
\end{align}
where the sole difference compared with (\ref{eq: CRLB sigma beta}) is the set of $ n $'s for the summation. 

Jointly inspecting the two CRLBs, we see that they only differ by a linear coefficient; namely,
\begin{align}
	\frac{\myCRB{\hat{\sigma}_{\beta}}}{\myCRB{\hat{\sigma}_{\alpha}}} = \frac{
		\sum_{n=\hat{n}}^{N-1}f[n]^2(\mu-\dx n)^4	
	}{\sum_{n=0}^{\hat{n}-1}f[n]^2(\mu-\dx n)^4 }.
\end{align}
\textit{An insight from this result is that we only need a linear combination of the two $ \sigma $ estimates obtained in (\ref{eq: sigma alpha sigma beta}) to achieve an asymptotically optimal combined estimation.} To further illustrate this, let us consider the following linear combination,
\begin{align} \label{eq: sigma = rho sigma alpha + (1-rho) sigma beta}
	\hat{\sigma} = \rho\hat{\sigma}_{\alpha} + (1-\rho) \hat{\sigma}_{\beta},~\rho\in(0,1). 
\end{align}
The mean squared error (MSE) of the combined estimate can be computed as 
\begin{align} \label{eq: MSE signal estimation}
	& \myExp{\myBrktRound{\hat{\sigma}-\sigma}^2} = \myExp{\myBrktRound{\rho\hat{\sigma}_{\alpha} + (1-\rho) \hat{\sigma}_{\beta}-\rho\sigma - (1-\rho)\sigma}^2}\nonumber\\
	& = \rho^2 \myExp{\myBrktRound{\hat{\sigma}_{\alpha}-\sigma}^2} + (1-\rho)^2  \myExp{\myBrktRound{\hat{\sigma}_{\beta}-\sigma}^2}\nonumber\\
	& \sim \rho^2  \myCRB{\hat{\sigma}_{\alpha}} + (1-\rho)^2  \myCRB{\hat{\sigma}_{\beta}} ,
\end{align}
where ``$ a \sim b  $" denotes that $ a $ asymptotically approaches $ b $ or a constant linear scaling of $ b $.

Solving $ \frac{\partial \myExp{\myBrktRound{\hat{\sigma}-\sigma}^2}}{\partial \rho}=0 $ leads to the following optimal $ \rho $,
\begin{align}%
	 \rhostar &= {\frac{
		\myCRB{\hat{\sigma}_{\beta}}
	}{\myCRB{\hat{\sigma}_{\beta}} + \myCRB{\hat{\sigma}_{\alpha}} }}
=
	\frac{
		\sum_{n=\hat{n}}^{N-1} f[n]^2(\mu-\dx n)^4	
	}{ 
		\sum_{n=0}^{N-1}f[n]^2(\mu-\dx n)^4 
	} \nonumber
\end{align}
where the CRLB expressions (\ref{eq: CRLB sigma beta}) and (\ref{eq: CRLB sigma alpha}) have been applied.
The optimality of $ \rhostar $ can be validated by plugging $ \rho=\rhostar $ into (\ref{eq: MSE signal estimation}). Doing so yields
\begin{align} \label{eq: MSE = CRLB optimal rho}
	\myExp{\myBrktRound{\hat{\sigma}-\sigma}^2}\sim \frac{\sigma_{\xi}^2}{\frac{\sum_{n=0}^{N-1}f[n]^2(\mu-\dx n)^4	}{\sigma^6} }.
\end{align}
From the index ranges of the summations in (\ref{eq: CRLB sigma beta}), (\ref{eq: CRLB sigma alpha}) and (\ref{eq: MSE = CRLB optimal rho}), we can see that the right-hand side of (\ref{eq: MSE = CRLB optimal rho}) becomes the CRLB of the $ \sigma $ estimation that is obtained based on all samples at hand --- \textit{the best estimation performance for any unbiased estimator of $ \sigma $ based on $ y[0], y[1],\cdots,y[N-1]$}. That is, taking $ \rho=\rhostar $ in (\ref{eq: sigma = rho sigma alpha + (1-rho) sigma beta}) leads to an asymptotically optimal unbiased $ \sigma $ estimation. 

Note that $ f[n] $ used for calculating $ \rhostar $ is unavailable. Thus, we replace $ f[n] $ with its noisy version $ y[n] $, attaining the following practically usable coefficient
\begin{align} \label{eq: rho optimal with y[n]}
\rhostar	\approxeq \frac{
		\sum_{n=\hat{n}}^{N-1} y[n]^2(\mu-\dx n)^4	
	}{ 
		\sum_{n=0}^{N-1} y[n]^2(\mu-\dx n)^4 
	}.
\end{align}
If we define $ A^2/\sigma_{\xi}^2 $ as the estimation signal-to-noise ratio (SNR), where $ \sigma_{\xi}^2 $ is the power of the noise term $ \xi[n] $ in (\ref{eq: y[n]}), then the equality in (\ref{eq: rho optimal with y[n]}) can be approached, as the estimation SNR increases.

\begin{table*}[b!]
	\hrule
	\centering
	\caption{A summary of the simulated methods, where the running time of each method is averaged over $ 10^5 $ independent trials. The simulations are run in MATLAB R2021a installed on a computing platform equipped with the Intel Xeon Gold 6238R 2.2GHz 38.5MB L3 Cache (Maximum Turbo Frequency 4.0GHz, Minimum 3.0GHz).
		\label{tab: code names and steps}}
	\begin{tabularx}{18cm}{c | c | p{12cm} | c }
		
		\hline
		
		Code Name
		& Method
		& Fitting Steps
		& Time ($ \mu $s)
		\\ 
		\hline
		
		M1
		& \cite{GF_Nahhal_SPM2019}  
		&  Estimate $ \hat{A} $ and  $ \hat{n} $ as done in (\ref{eq: max y[n] for A and n estimates}), where $ \hat{n} $ leads to $ \hat{\mu}=\hat{n}\dx $; estimate $ \hat{\sigma} $ using (\ref{eq: sigma estimate nahhal}).
		& $ 76.79 $
		\\
		\hline
		
		M2
		& \cite{GF_Nahhal_SPM2019} \& \cite{GF_Guo_SPM2011}
		& \makecell[l]{\textbf{Stage 1}: Run M1 first, getting initial $ \hat{A} $, $ \hat{\mu} $ and $ \hat{\sigma} $; \\ 
			\noindent \textbf{Stage 2}: Perform the iterative WLS based on (\ref{eq: WLS}), where $ \mathbf{w}_0=\myBrktSqr{\hat{A}e^{-\frac{(n\dx-\hat{n}\dx)^2}{2\hat{\sigma}^2}}\Big|_{n=0,1,\cdots,N-1}}^{\mathrm{T}} $.
		} 
		& $ 275.00 $
		\\
		\hline
		
		M3
		& New
		& Estimate $ \hat{\mu} $ based on (\ref{eq: mu hat}); estimate $ \hat{A} $ using (\ref{eq: A hat first refinement}); perform the estimators in (\ref{eq: sigma alpha sigma beta}), attaining $ \sigma_{\alpha} $ and $ \sigma_{\beta} $; combine the two estimates in the linear manner depicted in (\ref{eq: sigma = rho sigma alpha + (1-rho) sigma beta}), where the optimal $ \rho $, as approximately calculated in (\ref{eq: rho optimal with y[n]}), is used as the combination coefficient; refine $ \hat{A} $ as done in (\ref{eq: A refine min mse}).
		& $ 304.61 $
		\\ 
		\hline
		
		M4
		& New \& \cite{GF_Guo_SPM2011}
		& \makecell[l]{\textbf{Stage 1}: Run M3 first, getting initial $ \hat{A} $, $ \hat{\mu} $ and $ \hat{\sigma} $; \\ 
			\noindent \textbf{Stage 2}: Same as in M2.}
		& $ 502.81 $
		\\
		\hline
		
		M5
		& \cite{GF_Guo_SPM2011}
		& Perform the iterative WLS based on (\ref{eq: WLS})
		& $ 1,009.72 $
		\\		
		\hline 
		
	\end{tabularx}
	
\end{table*}

\begin{table}[t!]
	\caption{Simulation parameters, where ``Var." stands for ``Variable", and $ \mathcal{U}[a,b] $ denotes the uniform distribution in $ [a,b] $.}
	\label{tab: simulation parameter}
	
	\begin{tabularx}{\linewidth}{c|p{5.2cm}|p{2.105cm}}
		\hline
		Var.
		& Description
		& Value
		\\
		\hline
		
		$ A $
		& A parameter of the Gaussian function, determining its maximum amplitude; see (\ref{eq: f(x)})
		& $ 1 $
		\\
		\hline
		
		$ \mu $
		& A parameter of the Gaussian function, indicating its peak location; see (\ref{eq: f(x)})
		& $ \mathcal{U}[8,9] $
		\\
		\hline
		
		$ \sigma $
		& A parameter of the Gaussian function, indicating its width of the principal region; see Fig. \ref{fig: WLS fitted Gaussian curves}(c)
		& $ \mathcal{U}[1,1.3] $
		\\
		\hline
		
		$ x $
		& Function variable 
		& [0,10]
		\\
		\hline
		
		$ \dx $
		& Sampling interval of $ x $; see (\ref{eq: y[n]})
		& 0.01
		\\
		\hline

		$ \mk $
		& Intermediate variable used in the proposed estimators given in (\ref{eq: sigma alpha sigma beta})
		& $ 0.1:0.01:10 $		
		\\
		\hline
		
		$ L $
		& Windows size for estimating $ \mu $ as done in (\ref{eq: mu hat})
		& $ 3 $
		\\
		\hline
		
		~
		& Number of iterations for M5
		& 12
		\\
		\hline
		
		~
		& Number of iterations in Stage 2 of M2/M4
		& 2
		\\
		\hline

		$ \sigma_{\xi}^2 $
		& Power of the additive noise $ \xi[n] $ given in (\ref{eq: y[n]})
		& $ -10:0.5:20 $ dB
		\\
		\hline 
	\end{tabularx}
	
\end{table}

\subsection{Improving Estimation Performance of $ A $ and $ \mu $} \label{subsec: improve A and mu estimates}

{So far, we have focused on introducing the novel estimation of $ \sigma $. From (\ref{eq: sigma alpha sigma beta}), we can see that the proposed $ \sigma $ estimation requires the estimations of $ A $ and $ \mu $, i.e., $ \hat{A} $ and $ \hat{n}\dx $ therein.} 
To ensure a clear logic flow, we used the naive way of estimating these two parameters, as described in (\ref{eq: max y[n] for A and n estimates}). 
Here, we illustrate some simple yet more accurate methods for estimating $ A $ and $ \mu $.

For $ \mu $ estimation, we introduce a local averaging to reduce the impact of noise. Define a rectangular window function 
as $ W[n] = \frac{1}{L} $ for $ n=0,1,\cdots,L-1 $ and $ W[n] = 0 $ for other $ n $'s. 
The local averaging can be performed by using the window function to filter the sampled Gaussian function, i.e., $ y[n] $ given in (\ref{eq: y[n]}). An improved $ \mu $ estimation can be achieved by searching for the peak of the filtered Gaussian function and then constructing using the peak index.
This is expressed as
\begin{align} \label{eq: mu hat}
	\hat{\mu} = \myBrktRound{\hat{n} + \myFloor{\frac{L}{2}}}\dx,~\mathrm{s.t.}~\hat{n}:\arg \max_n \sum_{l=0}^{L-1} W[l] y[n+l].
\end{align}
The offset $ \myFloor{\frac{L}{2}} $ is added because, in theory,
if the sum of continuous $ L $ samples is maximum, those samples would be centered around the peak of a Gaussian function. Based on (\ref{eq: mu hat}), we also obtain an estimate of $ A $, i.e.,
\begin{align} \label{eq: A hat first refinement}
	\hat{A} = y\myBrktSqr{\hat{n}+ \myFloor{\frac{L}{2}}}. 
\end{align}

Use the two estimates obtained above in the proposed $ \sigma $ estimators, as given in (\ref{eq: sigma alpha sigma beta}). Then combine the two estimates as done in (\ref{eq: sigma = rho sigma alpha + (1-rho) sigma beta}) with the optimal combining coefficient given in (\ref{eq: rho optimal with y[n]}). This results in the final $ \sigma $ estimate. 
Unlike $ \hat{\mu} $ and $ \hat{\sigma} $ obtained using multiple samples, $ \hat{A} $ given in (\ref{eq: A hat first refinement}) is based on a single sample and hence can suffer from a large estimation error. Noticing this, we suggest another refinement of $ \hat{A} $ through minimizing the MSE, which is  calculated as 
\[\frac{1}{N}\sum_{n=0}^{N-1} \myBrktRound{xe^{-\frac{(n\dx-\hat{\mu})^2}{2\hat{\sigma}^2}} -y[n] }^2,\]
with respect to $ x $. The solution to the minimization is an improved $ A $ estimate, as given by
\begin{align}\label{eq: A refine min mse}
	\hat{A} = \frac{\myBrktRound{\sum_{n=0}^{N-1} e^{-\frac{(n\dx-\hat{\mu})^2}{2\hat{\sigma}^2}} y[n] }}{\myBrktRound{ \sum_{n=0}^{N-1} \myBrktRound{e^{-\frac{(n\dx-\hat{\mu})^2}{2\hat{\sigma}^2}}}^2 }}.
\end{align}

We summarize the proposed Gaussian fitting method in Table \ref{tab: code names and steps} under the code name M3. 
As mentioned at the end of Section \ref{subsec: iterative WLS}, we locate our method as an initial-stage fitting. For the second stage, we perform the iterative WLS with the initial weighting vector, i.e., $ \mathbf{w}_0 $ in (\ref{eq: WLS}), constructed using our initial fitting results. This two-stage fitting is named M4 in Table \ref{tab: code names and steps}. We underline that due to the high quality of the proposed initialization, M4 can converge much faster than the original iterative WLS, named M5 in Table \ref{tab: code names and steps}. This will be validated shortly by simulation results. Also provided in the table is the $ \sigma $ estimation method reviewed in Section \ref{subsec: Nahhal's sigma estimation}, as named M1. 
Moreover, the combination of M1 and the iterative WLS is referred to as M2 in Table \ref{tab: code names and steps}.

\section{Simulation Results}

\begin{figure}[t!]
	\centering
	\includegraphics[width=8cm]{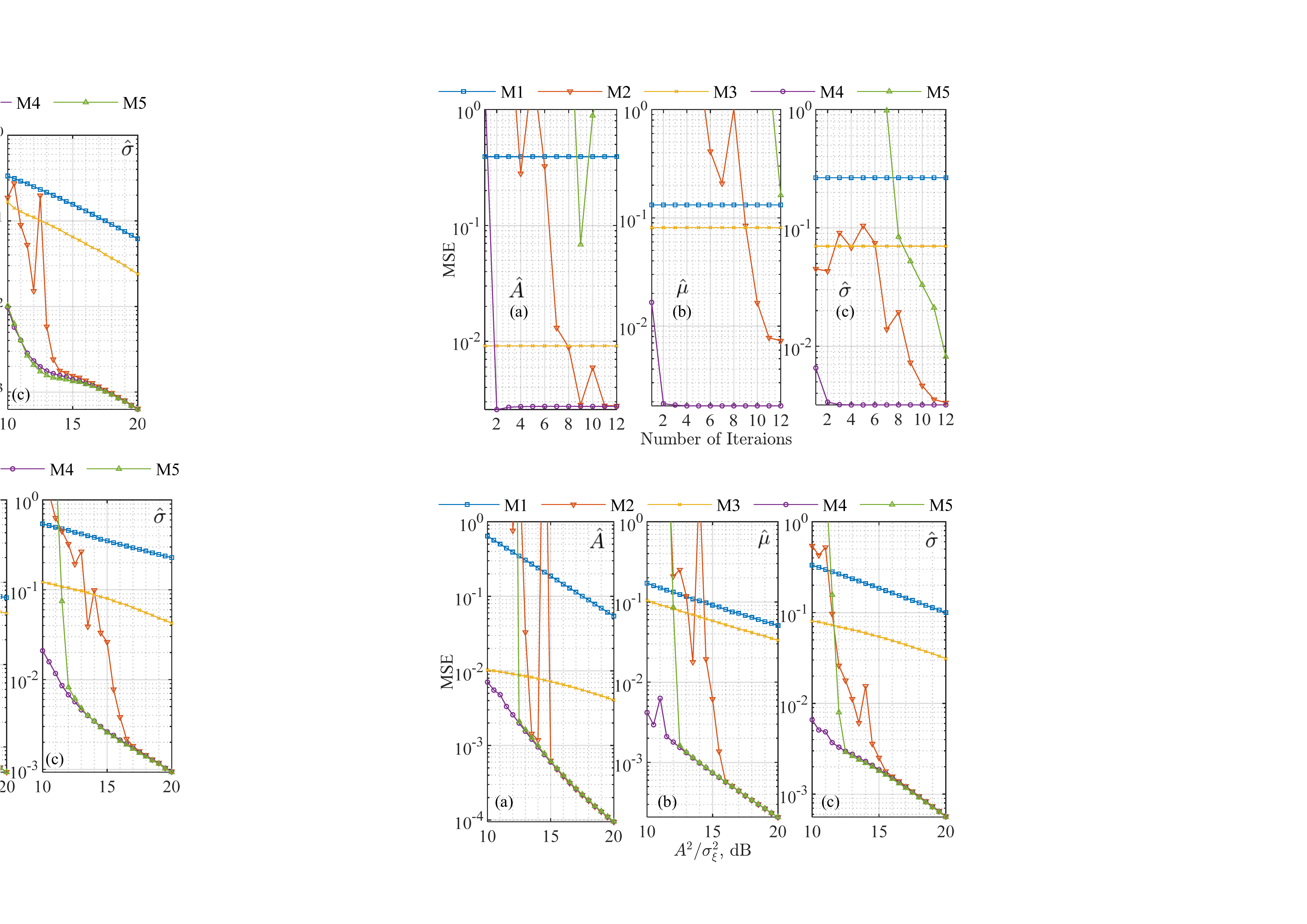}
	\caption{The MSEs of fitting results versus the SNR in the sampled Gaussian function with $ A=1 $, where $ \mu $ and $ \sigma $ are randomly generated based on the uniform distributions given in Table \ref{tab: simulation parameter}. Here, the sub-figures (a), (b) and (c) are for $ \hat{A} $, $ \hat{\mu} $ and $ \hat{\sigma} $, respectively. Note that $ \sigma_{\xi}^2 $ denotes the power of the noise term $ \xi[n] $ given in (\ref{eq: y[n]}). The MSE is calculated over $ 10^5 $ trials, each with independently generated and normally distributed noise.}
	\label{fig: mse mu9}
\end{figure}

\begin{figure}[t!]
	\centering
	\includegraphics[width=8cm]{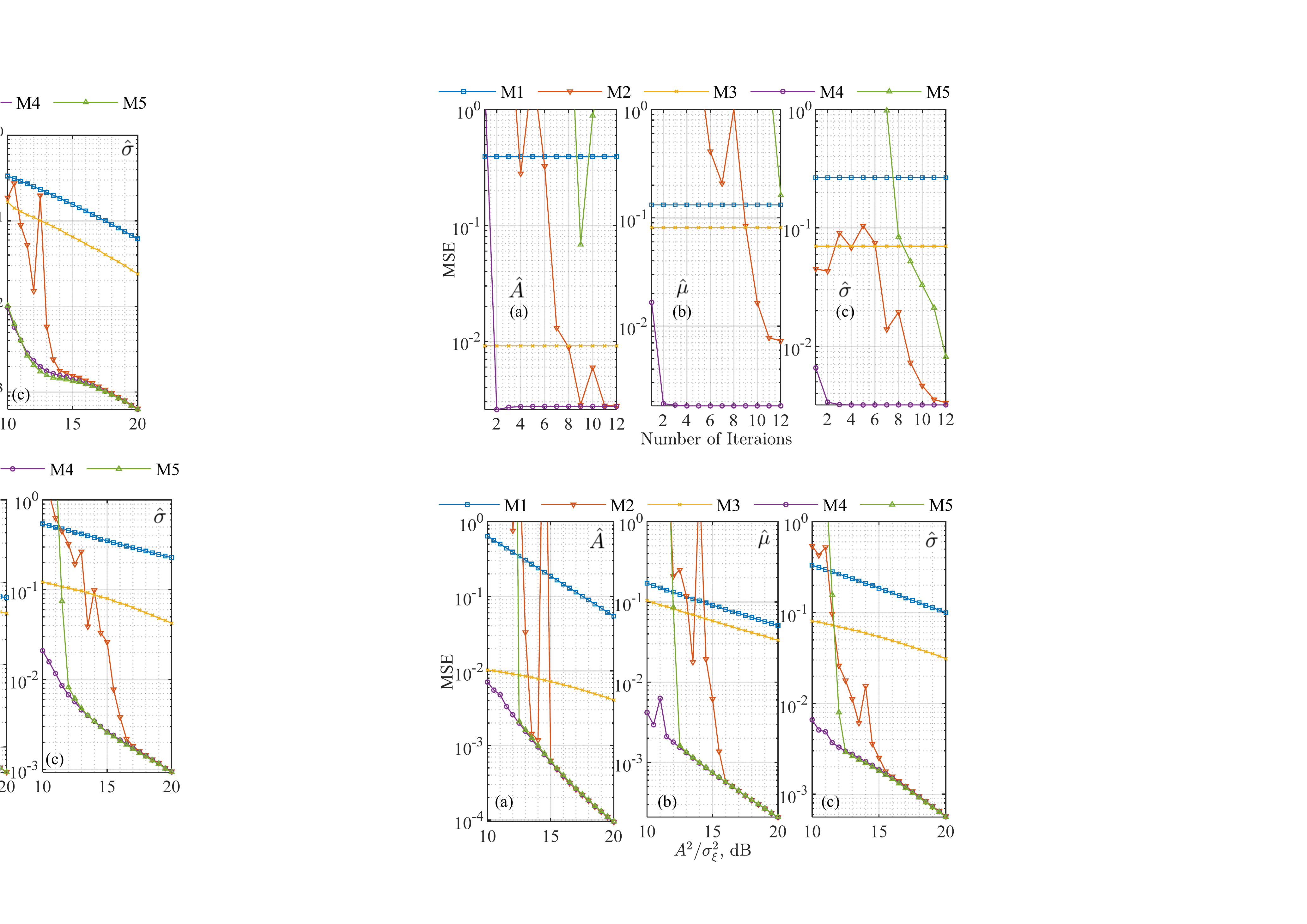}
	
	\caption{The MSEs of fitting results versus the number of iterations used in M5 and the second stage of M2/M4, where $ A=1 $, $ A^2/\sigma_{\xi}^2 $ is $ 12 $ dB, and $ \mu $ and $ \sigma $ are randomly generated based on the uniform distributions given in Table \ref{tab: simulation parameter}. The sub-figures (a), (b) and (c) are for $ \hat{A} $, $ \hat{\mu} $ and $ \hat{\sigma} $, respectively. The MSE is calculated over $ 10^5 $ trials, each with independently generated and normally distributed noise.}
	\label{fig: mse vs iter numbers}
\end{figure}

Simulation results\footnote{The MATLAB simulation codes for generating Figs. \ref{fig: mse mu9} and \ref{fig: mse vs iter numbers} can be downloaded from this link: \url{https://www.icloud.com/iclouddrive/005qLeEe1YOghnz4Om24ct76w\#publish_v2}
} 
are presented next to illustrate the performance of the five Gaussian fitting methods summarized in Table \ref{tab: code names and steps}.
Unless otherwise specified, the parameters summarized in Table \ref{tab: simulation parameter} are primarily used in our simulations. For the original iterative WLS, we perform $ 12 $ iterations so that it can achieve a similar asymptotic performance as M4 in the high-SNR region. (This will be seen shortly in Fig. \ref{fig: mse mu9}.) In contrast, when the initialization from either M1 or (the proposed) M3 is employed, only two iterations are performed for the iterative WLS algorithm. Note that the running time for each method is provided in Table \ref{tab: code names and steps}, {where each time result is averaged over $ 10^5 $ independent trials.}

Figs. \ref{fig: mse mu9} plots the MSEs of the five estimators listed in Table \ref{tab: code names and steps} against the estimation SNR, as given by $ A^2/\sigma_{\xi}^2 $. 
{Note that $ \mu $ and $ \sigma $ are randomly generated for the $ 10^5 $ independent trials, conforming to the uniform distributions given in Table \ref{tab: simulation parameter}. 
	As given in Table \ref{tab: simulation parameter}, $ x\in[0,10] $ is set in the simulation. Thus, the settings of $ \mu $ and $ \sigma $ make the Gaussian function to be fitted in each trial incompletely sampled with a long tail; a noisy version of the function is plotted in Fig. \ref{fig: WLS fitted Gaussian curves}(a). 
}

From Fig. \ref{fig: mse mu9}(b), we see that M3, which is based on the proposed local averaging in (\ref{eq: mu hat}), achieves an obviously better $ \mu $ estimation performance than M1, which is based on the naive method given in (\ref{eq: max y[n] for A and n estimates}). From Fig. \ref{fig: mse mu9}(c), we see that M3 substantially outperforms M1. This validates the competency of the proposed $ \sigma $ estimation scheme for the cases with the Gaussian function (to be fitted) incompletely sampled. 
From Fig. \ref{fig: mse mu9}(a), we see a significant improvement in M3, as compared with M1. 
This validates the advantage of using all samples for $ A $ estimation, as developed in (\ref{eq: A refine min mse}).

We remark that, as a price paid for performance improvement, the proposed initial fitting requires more computational time than M1; see the last column of Table \ref{tab: code names and steps} for comparison. 
However, it is noteworthy that, thanks to the signal processing tricks introduced in Section \ref{subsec: sigma estimation}, the proposed $ \sigma $ estimator, as established in (\ref{eq: sigma alpha sigma beta}), only involves simple floating-point arithmetic that can be readily handled by modern digital signal processors or field programming gate arrays.

From Fig. \ref{fig: mse mu9}, we further see that the proposed initial fitting results enable the iterative WLS to achieve much better performance for all three parameters than other initializations. The improvement is particularly noticeable in low SNR regions. 
Moreover, we underline that M4 based on the proposed initialization only runs two iterations, while the original iterative WLS, i.e., M5, runs $ 12 $
iterations. This, on the one hand, illustrates the critical importance of improving the initialization for the iterative WLS, a main motivation of this work. On the other hand, 
this validates our success in developing a high-quality initialization for the iterative WLS. 

{In spite of the random changing of $ \mu $ and $ \sigma $ over $ 10^5 $ independent trials, Fig. \ref{fig: mse mu9} shows that our proposed initial fitting enables the WLS to achieve consistently better and more stable performance than prior arts. 
This suggests that we have successfully relieved the dependence of the iterative WLS on the proportion of the tail region of a Gaussian function. In contrast, as illustrated in Section \ref{subsec: iterative WLS} and \ref{subsec: Nahhal's sigma estimation}, the performance of M5 and M1 can be subject to how complete a Gaussian function is sampled. M2, which is based on M1, also has the dependence.}

{Fig. \ref{fig: mse vs iter numbers} shows the MSEs of the five estimators listed in Table \ref{tab: code names and steps} against the number of iterations of the iterative WLS, performed in the second stage of M2/M4 and M5. For all three parameters, we can see that the proposed initialization (M3) non-trivially outperforms the state-of-the-art M1. We also see that M4 approximately converges after two iterations, while M2 and M5 present much slower convergence with the MSE performance inferior to M4 even after $ 12 $ iterations. These observations highlight the critical importance of a good initialization to the iterative WLS, particularly when the sampled Gaussian function is noisy and incomplete with a long tail. It again validates the effectiveness the proposed techniques, as enabled by the unveiled signal processing tricks, in these challenging scenarios. }

\section{Conclusions}
\label{secconc}

In this article, we develop a high-quality initialization method for the iterative WLS-based linear Gaussian fitting algorithm. 
This is achieved by a few signal processing tricks, as summarized below. 
\begin{enumerate}
\item We introduce a simple local averaging technique that reduces the noise impact on estimating the peak location of a Gaussian function, i.e., $ \mu $;

\item We provide a more precise integral result that is approximated by the summation of the Gaussian function samples, which results in two estimates of $ \sigma $;

\item We unveil the linear relation between the asymptotic performance of the two estimates and then design an asymptotically optimal combination of these estimates;

\item We also improve the estimation of the peak amplitude of a Gaussian function by minimizing the mean squared error of the initial Gaussian fitting.  
\end{enumerate}
Corroborated by simulation results, 
the proposed initialization can substantially improve the accuracy of the iterative WLS-based linear Gaussian fitting, even in challenging scenarios with strong noises and the incompletely sampled Gaussian function with a long tail. 
Notably, the performance improvement is also accompanied by improved fitting efficiency. 

\section{Acknowledgment}
{We thank Prof Rodrigo Capobianco Guido, \textit{IEEE
Signal Processing Magazine's Area Editor for
Columns and Forum} and Dr. Wei Hu, \textit{Associate Editor for Columns and Forum} for managing the review of our article. We also thank the editors and the 
anonymous reviewers for providing constructive
suggestions to improve our work.
We further acknowledge the support of the
Australian Research Council under
the Discovery Project Grant DP210101411.}

\section{Author}
\label{secaut}
\noindent \textbf{Kai Wu} (Kai.Wu@uts.edu.au, Member, IEEE) received a PhD from Xidian University, China in 2019, and a PhD from the University of Technology Sydney (UTS), Australia in 2020. His Xidian-PhD won the \textit{Excellent PhD Thesis award 2019} from the Chinese Institute of Electronic Engineering (EE). 
His UTS-PhD was awarded for \textit{The Chancellor's List 2020}. He is now a research fellow at the Global Big Data Technologies Centre (GBDTC), UTS. His research interests include array signal processing, and its applications in radar and communications.

\vspace{2mm}

\noindent \textbf{J. Andrew Zhang} (Andrew.Zhang@uts.edu.au, Senior Member, IEEE) 
received the Ph.D. degree from the Australian National University, in 2004. Currently, Dr. Zhang is an Associate Professor in the School of Electrical and Data Engineering, UTS. He was a researcher with Data61, CSIRO, Australia from 2010 to 2016, the Networked Systems, NICTA, Australia from 2004 to 2010, and ZTE Corp., Nanjing, China from 1999 to 2001. Dr. Zhang's research interests are in the area of signal processing for wireless communications and sensing. He has published more than 200 papers in leading international Journals and conference proceedings, and has won 5 best paper awards. He is a recipient of CSIRO Chairman's Medal and the Australian Engineering Innovation Award in 2012 for exceptional research achievements in multi-gigabit wireless communications.

\vspace{2mm}

\noindent \textbf{Y. Jay Guo} (Jay.Guo@uts.edu.au, Fellow, IEEE) received a
PhD Degree from Xian Jiaotong University, Xi’an China, in
1987. Currently, he is a Distinguished Professor and the Director
of GBDTC at UTS, and the Technical Director of the New South Wales Connectivity Innovation Network Australia. His research interest includes
antennas, mm-wave and THz communications and sensing
systems as well as big data technologies. He has published
six books and over 600 research papers including 320 journal
papers, most of which are in IEEE Transactions, and he holds
26 patents. He is a Fellow of the Australian Academy of
Engineering and Technology, a Fellow of IEEE and a Fellow of
IET, and was a member of the College of Experts of Australian
Research Council (ARC, 2016-2018). He has won a number
of most prestigious Australian Engineering Excellence Awards
(2007, 2012) and CSIRO Chairman’s Medal (2007, 2012). He
was named one of the most influential engineers in Australia
in 2014 and 2015, and one of the top researchers
in Australia in 2020 and 2021,respectively.

\bibliographystyle{IEEEtran}
\bibliography{IEEEabrv,bib_GF.bib}

\begin{thebibliography}{1}
\providecommand{\url}[1]{#1}
\csname url@samestyle\endcsname
\providecommand{\newblock}{\relax}
\providecommand{\bibinfo}[2]{#2}
\providecommand{\BIBentrySTDinterwordspacing}{\spaceskip=0pt\relax}
\providecommand{\BIBentryALTinterwordstretchfactor}{4}
\providecommand{\BIBentryALTinterwordspacing}{\spaceskip=\fontdimen2\font plus
\BIBentryALTinterwordstretchfactor\fontdimen3\font minus
  \fontdimen4\font\relax}
\providecommand{\BIBforeignlanguage}[2]{{%
\expandafter\ifx\csname l@#1\endcsname\relax
\typeout{** WARNING: IEEEtran.bst: No hyphenation pattern has been}%
\typeout{** loaded for the language `#1'. Using the pattern for}%
\typeout{** the default language instead.}%
\else
\language=\csname l@#1\endcsname
\fi
#2}}
\providecommand{\BIBdecl}{\relax}
\BIBdecl

\bibitem{book_formulasTables4SP_poularikas2018handbook}
A.~D. Poularikas, \emph{Handbook of formulas and tables for signal
  processing}.\hskip 1em plus 0.5em minus 0.4em\relax CRC press, 2018.

\bibitem{GF_applicationsWeb}
LogicPlum, ``What is a {Gaussian} function?''
  \url{https://logicplum.com/knowledge-base/gaussian-function/}, accessed:
  2022-03-11.

\bibitem{GF_SPletter2013}
E.~Kheirati~Roonizi, ``A new algorithm for fitting a {Gaussian} function riding
  on the polynomial background,'' \emph{IEEE Signal Process. Lett.}, vol.~20,
  no.~11, pp. 1062--1065, 2013.

\bibitem{GF_caruana1986fast}
R.~A. Caruana, R.~B. Searle, T.~Heller, and S.~I. Shupack, ``Fast algorithm for
  the resolution of spectra,'' \emph{Analytical chemistry}, vol.~58, no.~6, pp.
  1162--1167, 1986.

\bibitem{GF_Guo_SPM2011}
H.~Guo, ``A simple algorithm for fitting a {Gaussian} function [dsp tips and
  tricks],'' \emph{IEEE Signal Process. Mag.}, vol.~28, no.~5, pp. 134--137,
  2011.

\bibitem{GF_Nahhal_SPM2019}
I.~Al-Nahhal, O.~A. Dobre, E.~Basar, C.~Moloney, and S.~Ikki, ``A fast,
  accurate, and separable method for fitting a {Gaussian} function [tips and
  tricks],'' \emph{IEEE Signal Process. Mag.}, vol.~36, no.~6, pp. 157--163,
  2019.

\bibitem{book_kay1993fundamentals_estimation}
S.~M. Kay, \emph{Fundamentals of statistical signal processing: estimation
  theory}.\hskip 1em plus 0.5em minus 0.4em\relax Prentice-Hall, Inc., 1993.

\bibitem{Matlab_helpCenter}
MathWorks, ``Help center,'' \url{https://au.mathworks.com/help/matlab/},
  accessed: 2022-03-11.

\end{thebibliography}

\end{document}